\documentclass[]{spie}  

\usepackage{amsmath,amsfonts,amssymb}
\usepackage{graphicx}
\usepackage[colorlinks=true, allcolors=blue]{hyperref}

\usepackage{aas_macros}

\title{The Slicer Combined with Array of Lenslets for Exoplanet Spectroscopy (SCALES): driving science cases and expected outcomes}

\author[a]{Steph Sallum}
\author[b]{Andrew Skemer}
\author[c]{Deno Stelter}
\author[d]{Ravinder Banyal}
\author[b]{Natalie Batalha}
\author[e]{Natasha Batalha}
\author[f]{Geoff Blake}
\author[g]{Tim Brandt}
\author[h]{Zack Briesemeister}
\author[f]{Katherine de Kleer}
\author[i]{Imke de Pater}
\author[a]{Aditi Desai}
\author[j]{Josh Eisner}
\author[k]{Wen-fai Fong}
\author[e]{Tom Greene}
\author[l]{Mitsuhiko Honda}
\author[b]{Rebecca Jensen-Clem}
\author[b]{Isabel Kain}
\author[k]{Charlie Kilpatrick}
\author[b]{Renate Kupke}
\author[a]{Mackenzie Lach}
\author[m]{Michael C. Liu}
\author[c]{Bruce Macintosh}
\author[a]{Raquel A. Martinez}
\author[f,n]{Dimitri Mawet}
\author[j]{Brittany Miles}
\author[o]{Caroline Morley}
\author[p]{Diana Powell}
\author[d]{Ramya Sethuram}
\author[q]{Patrick Sheehan}
\author[r]{Justin Spilker}
\author[s]{Jordan Stone}
\author[t]{Arun Surya}
\author[d]{Sivarani Thirupathi}
\author[u]{Athira Unni}
\author[j]{Kevin Wagner}
\author[v]{Yifan Zhou}

\affil[a]{University of California, Irvine, Irvine, CA USA}
\affil[b]{University of California, Santa Cruz, Santa Cruz, CA, USA}
\affil[c]{University of California Observatories, Santa Cruz, CA, USA}
\affil[d]{Indian Institute of Astrophysics, Bangalore, India}
\affil[e]{NASA Ames Research Center, Moffett Field, CA, USA}
\affil[f]{California Institute of Technology, Pasadena, CA, USA}
\affil[g]{University of California, Santa Barbara, Santa Barbara, CA, USA}
\affil[h]{NASA Goddard Space Flight Center, Greenbelt, MD, USA}
\affil[i]{University of California, Berkeley, Berkeley, CA, USA}
\affil[j]{University of Arizona, Tucson, AZ, USA}
\affil[k]{Northwestern University, Evanston, IL, USA}
\affil[l]{Okayama University of Science, Okayama, Japan}
\affil[m]{University of Hawai’i, Honolulu, HI, USA}
\affil[n]{Jet Propulsion Laboratory, Pasadena, CA, USA}
\affil[o]{The University of Texas at Austin, Austin, TX, USA}
\affil[p]{The University of Chicago, Chicago, IL, USA}
\affil[q]{National Radio Astronomy Observatory, Charlottesville, VA, USA}
\affil[r]{Texas A\&M University, College Station, TX, USA}
\affil[s]{Naval Research Laboratory, Washington, DC, USA}
\affil[t]{Tata Institute of Fundamental Research, Mumbai, India}
\affil[u]{Aryabhatta Research Institute of Observational Sciences (ARIES), Nainital, India}
\affil[v]{The University of Virginia, Charlottesville, VA, USA}

\authorinfo{*Corresponding author information:\\ Steph Sallum\\E-mail: ssallum@uci.edu}

\begin{document} 
\maketitle

\begin{abstract}
The Slicer Combined with Array of Lenslets for Exoplanet Spectroscopy (SCALES) is a $2-5~\mu$m, high-contrast integral field spectrograph (IFS) currently being built for Keck Observatory. 
With both low ($R\lesssim250$) and medium ($R\sim3500-7000$) spectral resolution IFS modes, SCALES will detect and characterize significantly colder exoplanets than those accessible with near-infrared ($\sim1-2~\mu$m) high-contrast spectrographs.
This will lead to new progress in exoplanet atmospheric studies, including detailed characterization of benchmark systems that will advance the state of the art of atmospheric modeling. 
SCALES' unique modes, while designed specifically for direct exoplanet characterization, will enable a broader range of novel (exo)planetary observations as well as galactic and extragalactic studies. 
Here we present the science cases that drive the design of SCALES. 
We describe an end-to-end instrument simulator that we use to track requirements, and show simulations of expected science yields for each driving science case. 
We conclude with a discussion of preparations for early science when the instrument sees first light in $\sim2025$. 
\end{abstract}


\section{INTRODUCTION}
\label{sec:intro}  
The Slicer Combined with Array of Lenslets for Exoplanet Spectroscopy (SCALES)\cite{2022SPIE12184E..0IS} is a thermal-infrared, high-contrast integral field spectrograph (IFS) currently under construction for Keck Observatory. 
With a spectral range of $2-5~\mu$m, SCALES will be the first facility-class, high-contrast IFS to operate at these wavelengths. 
While SCALES is designed specifically to detect and characterize colder exoplanets than those accessible with near-infrared IFSs, it will enable a wide range of new science at Keck.
Its broader applications will include protoplanet characterization, protoplanetary disk mapping, and Solar System object monitoring, as well as characterization of supernova remnants, active galactic nuclei, and more.  

\begin{figure}[h!]
    \centering
    \includegraphics[width=\textwidth]{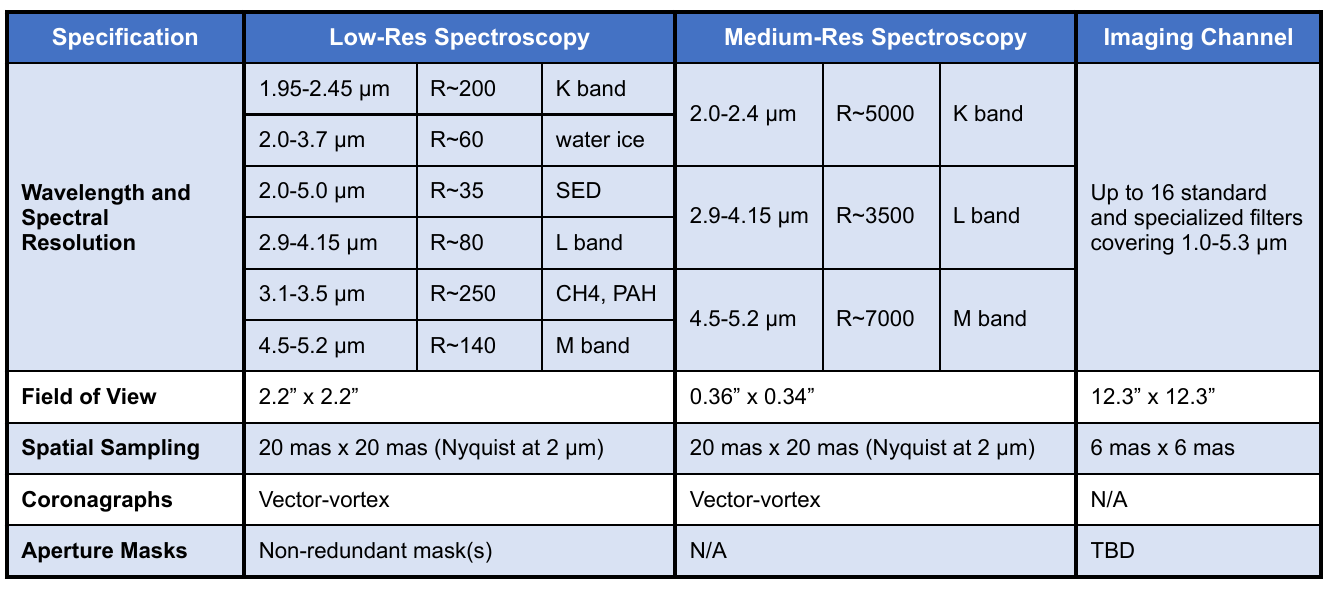}
    \caption{Summary of SCALES Top-Level Specifications}
    \label{fig:modes}
\end{figure}

SCALES will consist of a lenslet-based low-spectral-resolution IFS, a novel lenslet-plus-slicer (slenslit) medium-spectral-resolution IFS,\cite{2022SPIE12184E..45S} and an imaging channel.\cite{2022SPIE12188E..1UB} 
Figure \ref{fig:modes} summarizes the spectral resolutions, fields of view, spatial sampling, and high-contrast / high-resolution capabilities of each of SCALES' IFS and imaging modes. 
In addition to the baseline modes listed in Figure \ref{fig:modes}, the SCALES team is also exploring possible medium-resolution upgrades, with custom gratings and filters to enable higher spectral resolutions tailored for individual science cases (see Martinez \textit{et al.~}in these proceedings).\cite{martinez_inproc}

Here we describe the science cases that drive the SCALES design, as well as SCALES' expected science outcomes.
We first present the latest version of \texttt{scalessim}, which is an open-source end-to-end simulator that the SCALES team uses to track requirements and generate realistic mock observations. 
We then discuss SCALES' science objectives and anticipated impact for exoplanet detection and characterization, protoplanet detection and characterization, protoplanetary disk mapping, Solar System monitoring, and other applications beyond exoplanets and planetary science. 
We conclude by discussing ongoing preparations for early SCALES science in 2025.

For additional information about SCALES, please see the following proceedings from this conference:
\begin{itemize}
\itemsep0em
    \item \textbf{End-to-end spectrograph modeling:} Paper 12680-6 (Surya et al.~2023)\cite{surya_inproc}
    \item \textbf{Custom medium-spectral-resolution modes:} Paper 12680-69 (Martinez et al.~2023)\cite{martinez_inproc}
    \item \textbf{Diamond-turned optics performance:} Paper 12680-73 (Kain et al.~2023)\cite{kain_inproc}
    \item \textbf{Optics testing:} Paper 12680-75 (Gangadharan et al.~2023)\cite{aditig_inproc}
    \item \textbf{Medium-spectral-resolution point-spread function deconvolution:} Paper 12680-76 (Desai et al.~2023)\cite{desai_inproc}
    \item \textbf{Aperture mask designs:} Paper 12680-77 (Lach et al.~2023)\cite{lach_inproc}
\end{itemize}

\section{SCALESSIM: AN END-TO-END INSTRUMENT SIMULATOR}\label{sec:scalessim}

SCALES' design has been informed by a realistic end-to-end simulation tool (\texttt{scalessim}).\cite{2020SPIE11447E..4ZB} 
We developed this python-based instrument simulator for use in tracking requirements and performing trade studies.
We have used \texttt{scalessim} to make design choices such as filter selection, and to set requirements such as the low-resolution prism rotation angle, the pointing stability of the medium-resolution mode piezo mirror\cite{stelter_inproc}, the imaging channel plate scale, and more.
We have also used it to generate realistic mock observations for representative science scenes, and have developed an information content approach to flowing science requirements down to instrument requirements. \cite{2021SPIE11823E..08B} 
The \texttt{scalessim} package is publicly available on github\footnote{\url{https://github.com/scalessim/scalessim}} so that potential users can apply it to explore SCALES' science outcomes for various applications. 

The \texttt{scalessim} software takes as input either a target spectrum or a scene cube consisting of images at a range of wavelengths. 
In the former case, \texttt{scalessim} scales a set of Keck point-spread functions (PSFs) at a range of wavelengths by the flux values given in the input spectrum to create a scene cube.
In the latter, it convolves the 3d scene cube with the same set of PSFs.

Once a PSF-convolved scene has been generated, \texttt{scalessim} accounts for instrumental and astrophysical transmission in the following ways:
\begin{enumerate}
    \item \textbf{Sky transmission and emission:} \texttt{scalessim} accounts for sky transmission and emission using Mauna Kea sky background files provided by Gemini.\cite{1992nstc.rept.....L} These are selectable by the user according to airmass and precipitable water vapor. Martinez et al.~(2023)\cite{martinez_inproc} in these proceedings describes progress in providing a wider range of sky backgrounds for more realistic observation planning, and this feature will soon be incorporated into the main \texttt{scalessim} github repository. The scene flux at each wavelength is decreased by the sky transmission, and a uniform background is added to account for sky emission. 
    \item \textbf{Instrument transmission and emission:} \texttt{scalessim} next accounts for instrumental transmission and emission by decreasing the scene flux according to the telescope transmission, and then adding blackbody emission to account for the telescope plus adaptive optics backgrounds. The resulting scene flux is then decreased again to account for SCALES' instrumental transmission. All of the instrumental transmission and emission parameters are adjustable by the user. The default settings are a temperature of 285 K and emissivity of 0.4 for the telescope and AO system, and a constant SCALES throughput of 0.4. 
    \item \textbf{Filter curves:} \texttt{scalessim} includes realistic estimated filter curves for the planned filter set, as well as idealized top-hat filter curves (which assume perfect transmission over a user-specified wavelength range). It uses the expected curves by default, scaling the scene flux accordingly.
    \item \textbf{Detector quantum efficiency:} \texttt{scalessim} decreases the flux in the scene cubes to account for quantum efficiency. This is adjustable by the user. Since both SCALES detectors are NIRSpec flight spares, the default value is taken from Rauscher et al.~2014.\cite{2014PASP..126..739R} 
\end{enumerate}

\begin{figure}
    \centering
    \includegraphics[width=\textwidth]{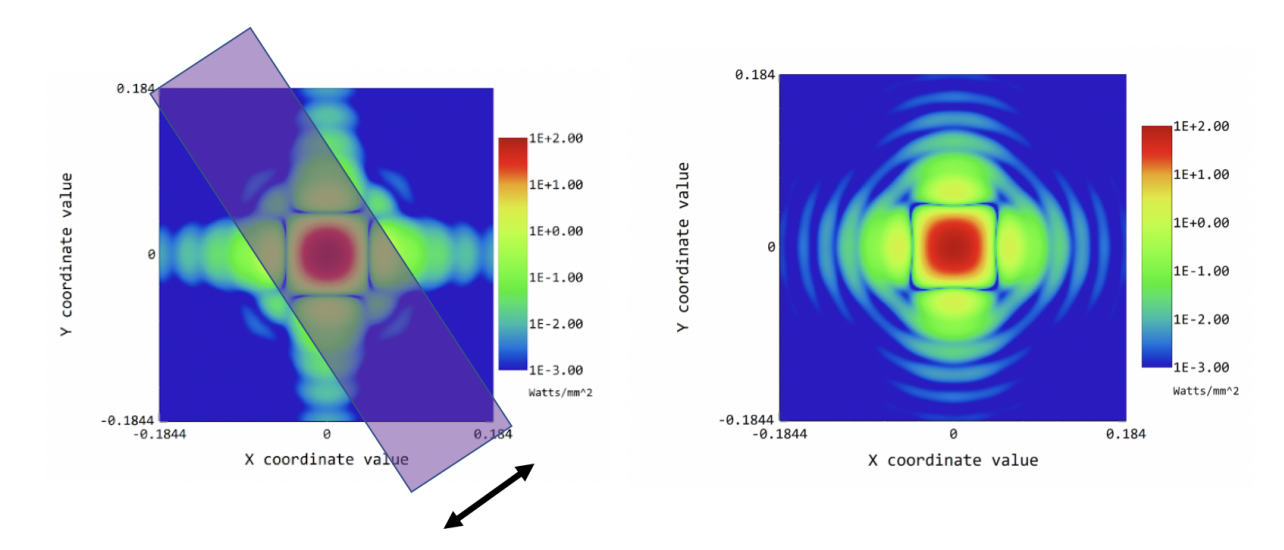}
    \caption{ 
        Left: The colorscale shows a 5.0 $\mu$m square lenslet PSF, with only diffraction by the lenslet taken into account. 
        The purple box shows  the the dispersion direction for SCALES' low-resolution IFS modes. 
        Right: The colorscale shows a 5.0 $\mu$m square lenslet PSF, with diffraction by the lenslet taken into account and the pinhole placed at the geometric focus of the lens. The pinhole throughput at 5 $\mu$m is $\sim$83\%.}
    \label{fig:lensletpsfs}
\end{figure}

After scaling the scene appropriately and adding backgrounds, \texttt{scalessim} converts from flux values to counts using a gain of 8 $\mathrm{e^-}$/DN.
It then uses monochromatic lenslet PSFs to create spectral traces on the detector. 
It convolves the scene flux with a lenslet PSF for each wavelength and spatial location in the scene.
These lenslet PSFs (Figure \ref{fig:lensletpsfs}) were generated with \texttt{poppy}\cite{2012SPIE.8442E..3DP} using a physical optics model of the lenslets plus a pinhole mask (which will be installed to suppress lenslet diffraction spikes and minimize optical crosstalk).
The monochromatic lenslet PSFs are then placed on the detector according to the expected dispersion curves for each prism / grating. 

As the last step in raw image generation, \texttt{scalessim} adds random and systematic noise sources. 
These include the following
\begin{enumerate}
    \item \textbf{Poisson noise:} \texttt{scalessim} uses the flux values in the raw, noiseless image to add Poisson noise contributions from the target and thermal background. 
    \item \textbf{H2RG detector noise:} This includes readout, pedestal, correlated and uncorrelated pink noise, alternating column noise, and ``picture frame" noise. \texttt{scalessim} adds these to the raw images following the prescription in Rauscher et al.~2015.\cite{2015PASP..127.1144R} 
    \item \textbf{Detector crosstalk:} \texttt{scalessim} adds appropriate levels of crosstalk for an H2RG following the implementation in George et al.~2018.\cite{2018arXiv180800790G}
\end{enumerate}

With raw images in hand, users can apply quick-look data reduction software to extract a cube of images at a range of wavelengths. 
We developed this bare-bones pipeline using simulated lenslet PSFs, prism dispersion curves, and \texttt{scalessim} data products.
We use the lenslet PSFs and dispersion curves to simulate calibration unit data, which for each wavelength consists of a set of monochromatic lenslet spots on the detector. 
We use these spots to generate a sparse matrix which, when multiplied by a flattened raw image, performs simple aperture photometry at the location of each lenslet spot for each wavelength. 
For now, this step sums all flux values contained within the region of the lenslet PSF that is within 25\% of its peak value, but adjusting this value will be explored more thoroughly in the future. 
The resulting array is then reshaped into a cube of images as a function of wavelength, with one pixel per lenslet.

\begin{figure}
    \centering
    \includegraphics{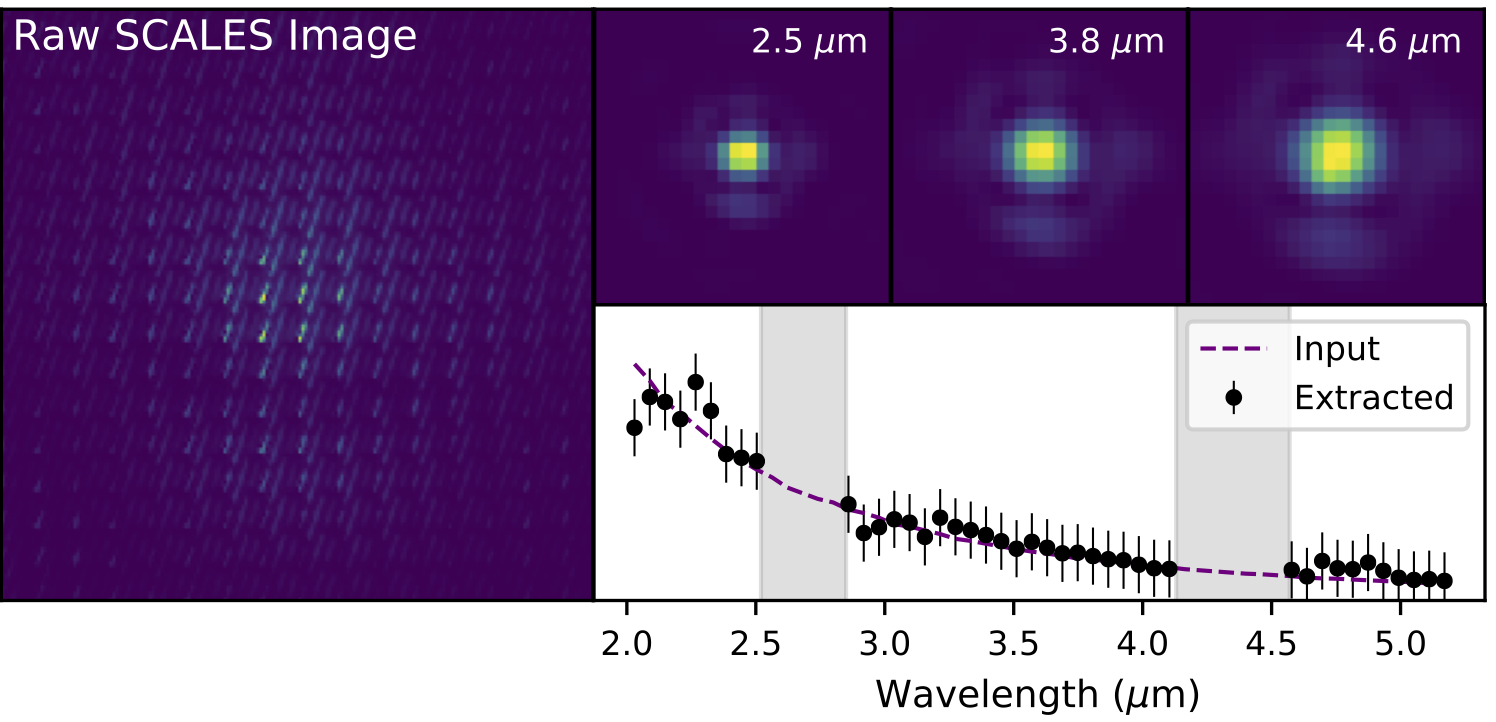}
    \caption[Simulated SCALES image from 1 second on an A0 star.]{
        Left: Simulated raw SCALES image of an A0 star (assuming 1 second of exposure), showing individual spectral traces (diagonal lines) superimposed on the Keck PSF (outer envelope). 
        Right: Top panels show extracted images from the bare-bones pipeline described in Section \ref{sec:scalessim}, and the bottom panel shows the extracted spectrum in arbitrary units (scattered points) plotted over the input (dashed line). Grey shading indicates essentially zero atmospheric transmission.}
    \label{fig:drp_example}
\end{figure}

This rectification process is similar to that for OSIRIS' quick look tool,\footnote{\url{https://www2.keck.hawaii.edu/inst/osiris/technical/recmat/}} which has associated overheads that are limited by file input/output, rather than the rectification step itself. 
We have confirmed that this is also case for the SCALES quick look pipeline.
The matrix multiplication step runs in approximately 30 ms, while the combined file input/output for the raw and cubified images and rectification matrix itself takes approximately 300 ms. 
The combined speed meets our quick-look requirement of producing a processed image cube in $<1$s, in order to allow for coronagraph alignment during observing. 
Figure \ref{fig:drp_example} shows an example cropped raw SCALES image of an unresolved star, along with example images and spectra extracted by the quick-look pipeline.

In addition to the quick-look pipeline, \texttt{scalessim} can be run to skip the raw frame generation step and instead produce 3D datacubes that assume ``perfect" spectral extraction. 
In this case, all of the incident flux on each lenslet (after adjusting for the Poisson noise sources described above) is enclosed in a single pixel in the 3D datacubes. 
Currently, when the extraction step is skipped in this mode, H2RG systematics and optical crosstalk are not included in the 3D extracted datacube.
We will improve this in future versions of \texttt{scalessim}.

\section{SCALES SCIENCE OBJECTIVES,  TRACEABILITY, AND EXPECTED OUTCOMES}

\begin{figure}
    \centering
    \includegraphics{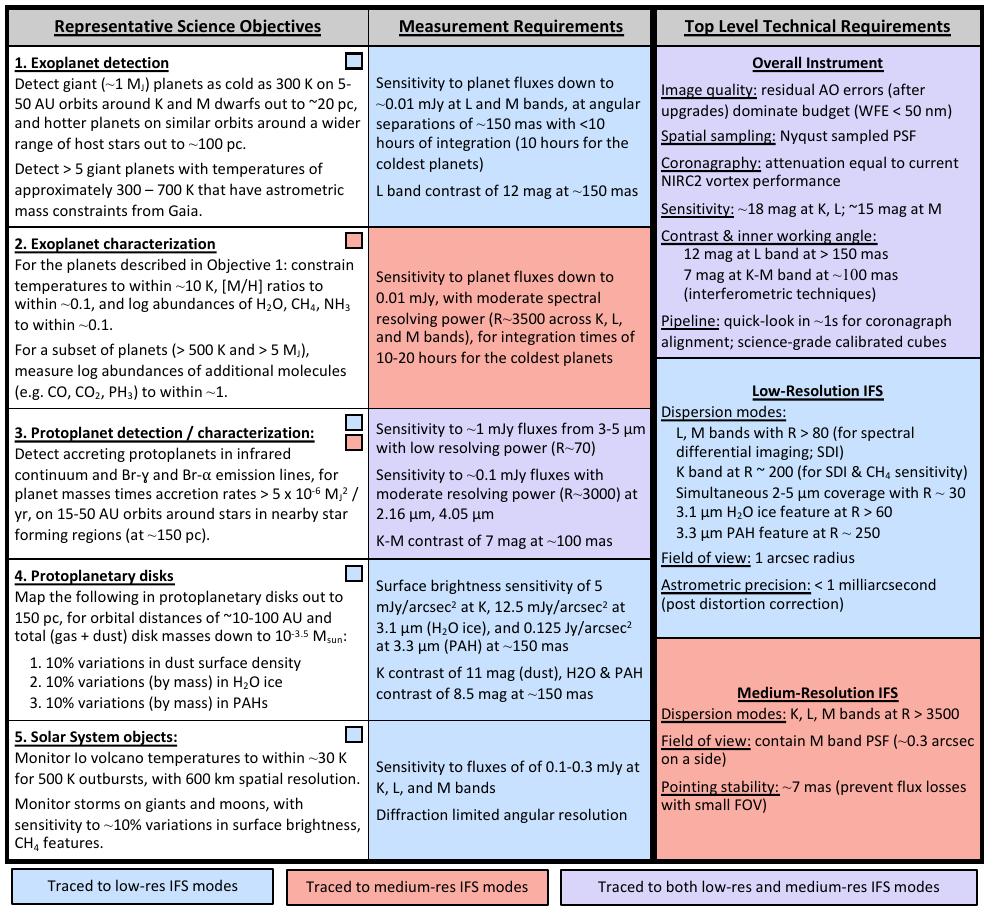}
    \caption{SCALES Science Traceability Matrix. These are representative science cases that are part of a larger science traceability matrix maintained by the SCALES team. In addition to the (exo)planetary cases above, we also track requirements for applications such as transients, galaxies, and the galactic center.}
    \label{fig:traceability}
\end{figure}

Here we describe the driving science cases that set the technical requirements for the SCALES modes shown in Figure \ref{fig:modes}.
The top-level science objectives for these cases are summarized in Figure \ref{fig:traceability}, along with their corresponding measurement requirements and top-level technical requirements. 
In Figure \ref{fig:traceability} and in the subsections that follow we focus primarily on the low- and medium-resolution IFS science requirements, but we note that the SCALES team also tracks science requirements for the imaging channel.
In each of the following subsections we highlight some of SCALES' anticipated science outcomes.

\subsection{Exoplanet Detection and Characterization}

\begin{figure}
    \centering
    \includegraphics[width=0.95\textwidth]{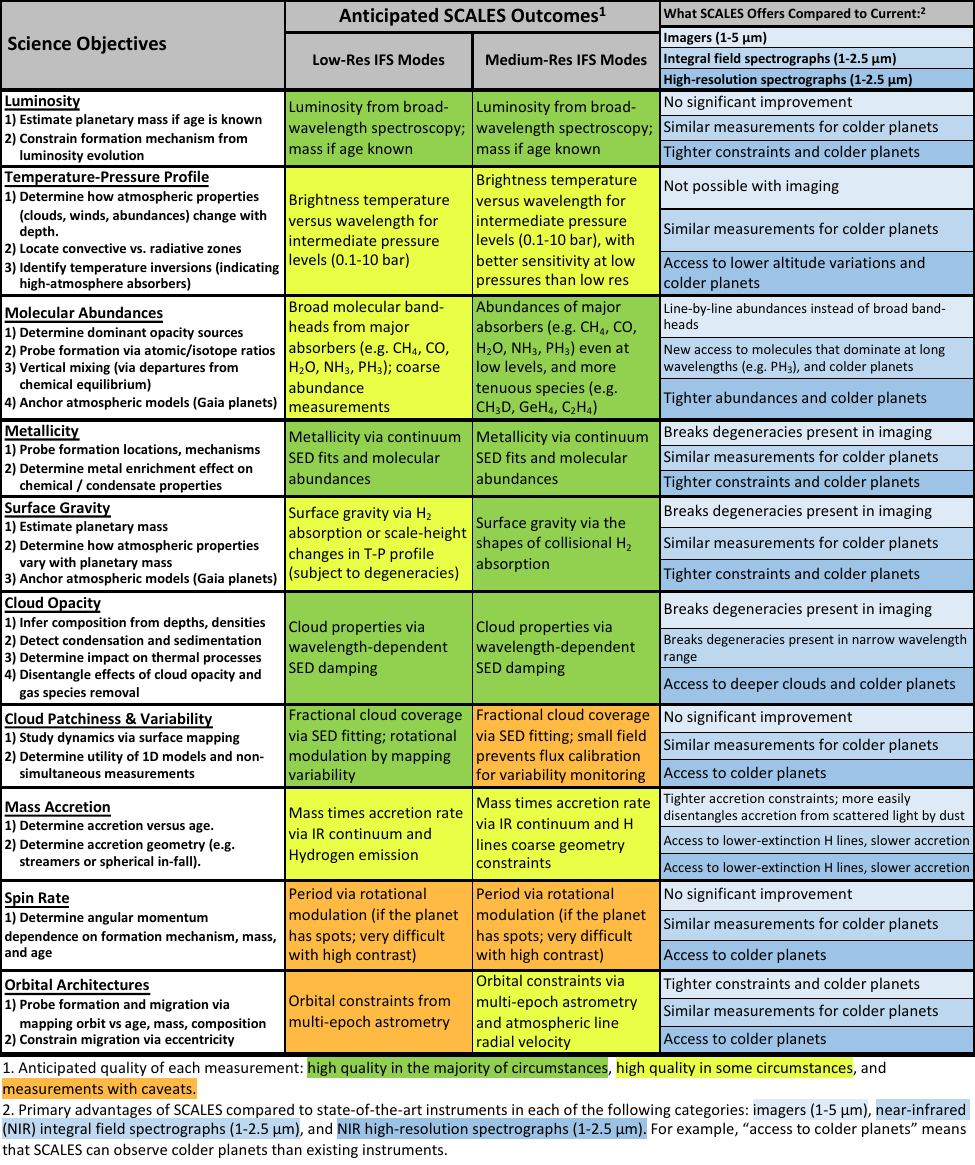}
    \caption{Summary of exoplanet characterization studies enabled by SCALES' low-resolution and medium-resolution IFS modes. The first column describes the main science objectives for a variety of exoplanet characterization studies. The second and third columns describe measurements to be made by the low-resolution and medium-resolution IFS modes, respectively. In the second and third columns, the color indicates the anticipated quality of each measurement, with green representing those that will be consistently high-quality, yellow showing those that will be high-quality in some circumstances, and orange indicating measurements with caveats. The fourth column describes how SCALES will improve upon current measurements with imagers (light blue), near-infrared IFSs (medium blue), and high-resolution spectrographs (dark blue).}
    \label{fig:characterization}
\end{figure}

The incomplete census of directly characterized exoplanets has limited our understanding of atmospheric physics and evolution.
SCALES is designed to make progress in this area by detecting colder exoplanets than those accessible with current, near-infrared integral field spectrographs. 
This will enable a wide range of new science outcomes, which are summarized in Figure \ref{fig:characterization}. 
SCALES' low-resolution IFS modes will enable robust measurements of planet luminosities, metallicities, and clouds. 
The medium-resolution modes will add measurements of atmospheric abundances, surface gravities, and more. 
Here we describe some of SCALES' premiere exoplanet science applications, including their top-level requirements and SCALES' expected yields. 

\subsubsection{Detecting Cold Exoplanets at Low Spectral Resolution}

\begin{figure}
    \centering
    \includegraphics[width=\textwidth]{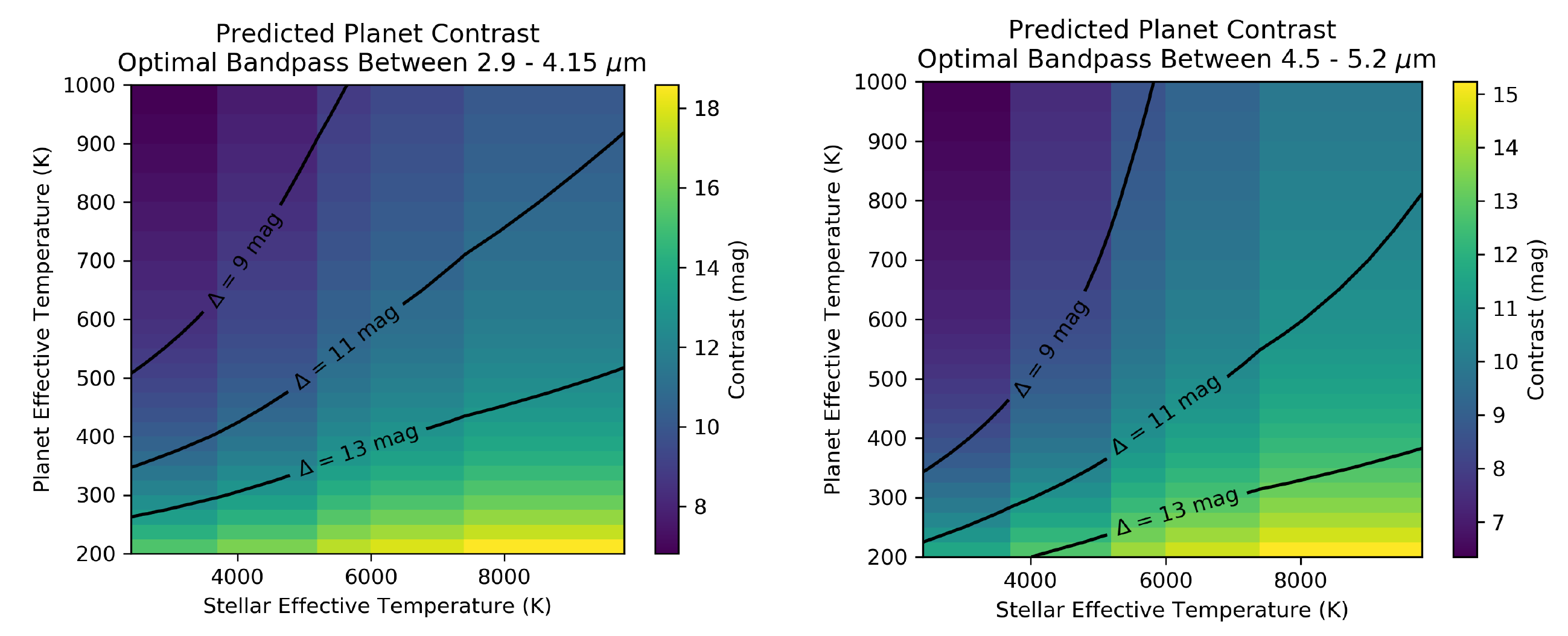}
    \caption{Expected planet contrasts used to set SCALES' top-level contrast requirements. In both panels, the colorscale shows predicted planet contrasts for a 1 R$_\mathrm{J}$ planet (assuming Sonora\cite{2021ApJ...920...85M} atmospheric models) as a function of effective temperature and stellar host temperature (assuming PHOENIX\cite{2010ascl.soft10056B} models). Left and right panels, respectively, show expectations for optimal bandpasses calculated for the SCALES L and M band low-resolution modes. In both panels the black lines indicate contrast levels of 9-13 magnitudes, which are achievable with current Keck adaptive optics.\cite{2019AJ....157...33M}}
    \label{fig:optimal}
\end{figure}

The coldest directly-imaged exoplanet to date (51 Eri b) has an effective temperature of $\sim650-700$ K.\cite{2015Sci...350...64M}
SCALES' top-level exoplanet detection goal is to detect giant ($\sim1~\mathrm{M_J}$) planets with temperatures down to $\sim300$ K, on orbits of $\sim5-50$ AU around K and M dwarfs out to $\lesssim20$ pc (and hotter planets around a wider range of stars; Figure \ref{fig:traceability}). 
Thanks to the wavelength coverage and spectral resolution of the low-resolution IFS, optimal bandpasses can be constructed to minimize the contrast of these exoplanets at thermal infrared wavelengths where they are already relatively bright (Figure \ref{fig:optimal}). 
SCALES will also leverage the performance of Keck adaptive optics to accomplish this goal, including a planned $\sim$3000 actuator deformable mirror upgrade that is part of the High-Order All Sky Keck Adaptive Optics (HAKA) project.\footnote{Expected to be online when SCALES sees first light: \url{https://www.keckobservatory.org/wp-content/uploads/2023/08/Keck2035_StratPlan_web.pdf}}

\begin{figure}
    \centering
    \includegraphics[width=\textwidth]{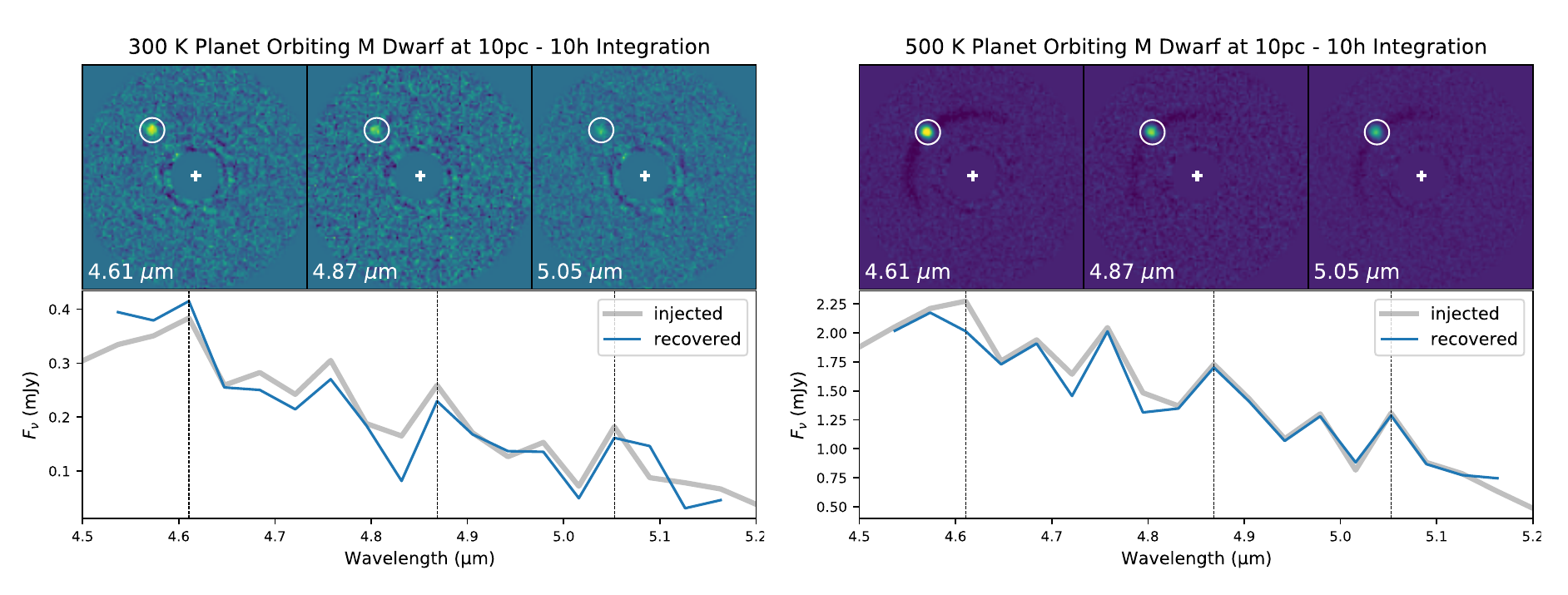}
    \caption{Example simulated SCALES low-resolution M band observations of a 300 K, 1 R$_\mathrm{J}$ planet (left), and a 500 K, 1 R$_\mathrm{J}$ planet (right), orbiting M dwarf host stars located at distances of 10 pc with semimajor axes of 6 AU. Both simulations use PSFs consistent with the predicted performance of Keck's High-Order All Sky Adaptive Optics (HAKA) upgrade. They also assume 10h of total integration time, 80$^\circ$ of parallactic angle evolution, and perfect telluric correction. In future work we will simulate these science cases with the realistic and variable precipitable water vapor settings described in Martinez et al.~(2023) in these proceedings.\cite{martinez_inproc}}
    \label{fig:coldplan}
\end{figure}

\begin{figure}
    \centering
    \includegraphics{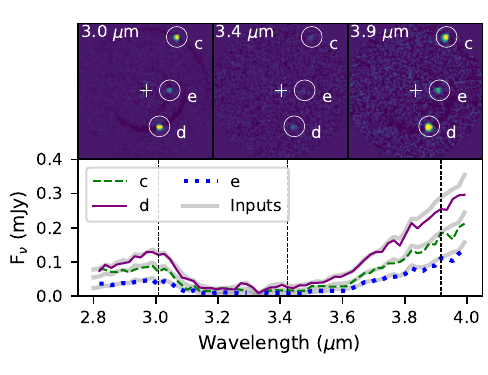}
    \caption{Example simulated SCALES L band observations of an HR 8799 analog, with three planets having effective temperatures between 900 K and 1100 K and radii of 1 R$_\mathrm{J}$. Top panels show ADI-processed slices from a low-resolution L band mock observation of a system of exoplanets similar to HR 8799. The bottom panel shows the recovered spectra for the three injected planets (green, purple, blue lines) plotted over the injected spectra (grey lines). The vertical lines in the bottom panel indicate the wavelengths of the three images in the top panels. The simulated observations assumed one hour of total integration time spread over $\sim80^\circ$ of parallactic angle evolution.}
    \label{fig:hr8799}
\end{figure}

Figure \ref{fig:coldplan} shows representative mock observations of cold planets in SCALES' low-resolution M band mode.
These simulations assume HAKA adaptive optics performance, with simulated HAKA point-spread functions being passed to \texttt{scalessim}.
They also assume a deep observation with 10 hours of total integration and $\sim80^\circ$ of parallactic angle evolution. 
They optimistically assume a static precipitable water vapor value and perfect telluric correction, and can thus be thought of as best-case results given the expected adaptive optics performance and thermal emission from the sky and instrument. 
We will explore the effects of varying conditions more realistically in future work by using the variable atmospheric settings described in Martinez et al.~(2023)\cite{martinez_inproc} in these proceedings. 
Meeting the contrast and sensitivity requirements for cold planet detection will mean that SCALES will be capable of low-resolution spectroscopy of warmer exoplanets around a wider variety of host stars in a shallower observation (Figure \ref{fig:hr8799}).

\subsubsection{Gaia-Selected Exoplanet Detections}\label{sec:gaia}
Approximately when SCALES comes online, the Gaia\cite{2016A&A...595A...1G} mission is expected to release a catalog of $\sim70,000$ exoplanets.\cite{2014ApJ...797...14P}
Some of this planet population will be uniquely accessible with SCALES thanks to its combination of high angular resolution, high contrast, and long wavelength coverage. 
Each one of these planets will be a valuable benchmark for atmospheric studies, since the dynamical mass measurement will break degeneracies typically present in atmospheric modeling.

We use \texttt{scalessim} to predict the number of astrometrically-detected planets that could be directly imaged with SCALES' low-resolution L-band IFS mode during a deep observation.
We assume the same simulated Gaia and Gaia plus WFIRST astrometrically-detected planet populations as those described in Brandt et al.~2019.\cite{2019BAAS...51c.269B}
These use a SAG13 underlying planet population with an exponential drop-off in giant planet occurrence rates.\cite{2019BAAS...51c.269B,2011PASP..123..692M}
We also assume a uniform distribution of planet ages between 1 Myr and 10 Gyr, and use COND cooling models\cite{2003A&A...402..701B} to predict effective temperatures based on planet ages and masses.
We then use spectra from Morley et al.~2014\cite{2014ApJ...787...78M} to determine both the planet photometry and the weighted IFS bandpass that maximizes the planet signal-to-noise ratio.

We compare the planet contrast and photometry to SCALES' achievable contrast and sensitivity, respectively to determine whether the planet is detectable. 
For current Keck natural guide star (NGS) adaptive optics performance, we begin with the Keck/NIRC2 contrast curve from Mawet et al.~2019\cite{2019AJ....157...33M}, scale its separation by 3.8/4.8 to convert from M to L$^\prime$ angular resolution, and assume a sensitivity floor of 18.8 mag for 10 hours of integration. 
We assume this performance for bright stars, and a linear contrast roll-off of $\sim1.5$ mag between 5th and 12th magnitude guide stars. 
We also predict planet yields for the planned HAKA upgrade.
We generate these contrast curves by using predicted HAKA point-spread functions as inputs for \texttt{scalessim}.
We then process the simulated SCALES+HAKA L-band images with VIP\cite{2017AJ....154....7G,2023JOSS....8.4774C} to produce an angular differential imaging\cite{2012ApJ...755L..28S} contrast curve. 
For both current NGS and HAKA, we assume a $\sim1$ mag contrast gain from additional spectral-differential imaging (SDI)\cite{2006SPIE.6272E..2DB} processing.

\begin{figure}
    \centering
    \includegraphics[width=\textwidth]{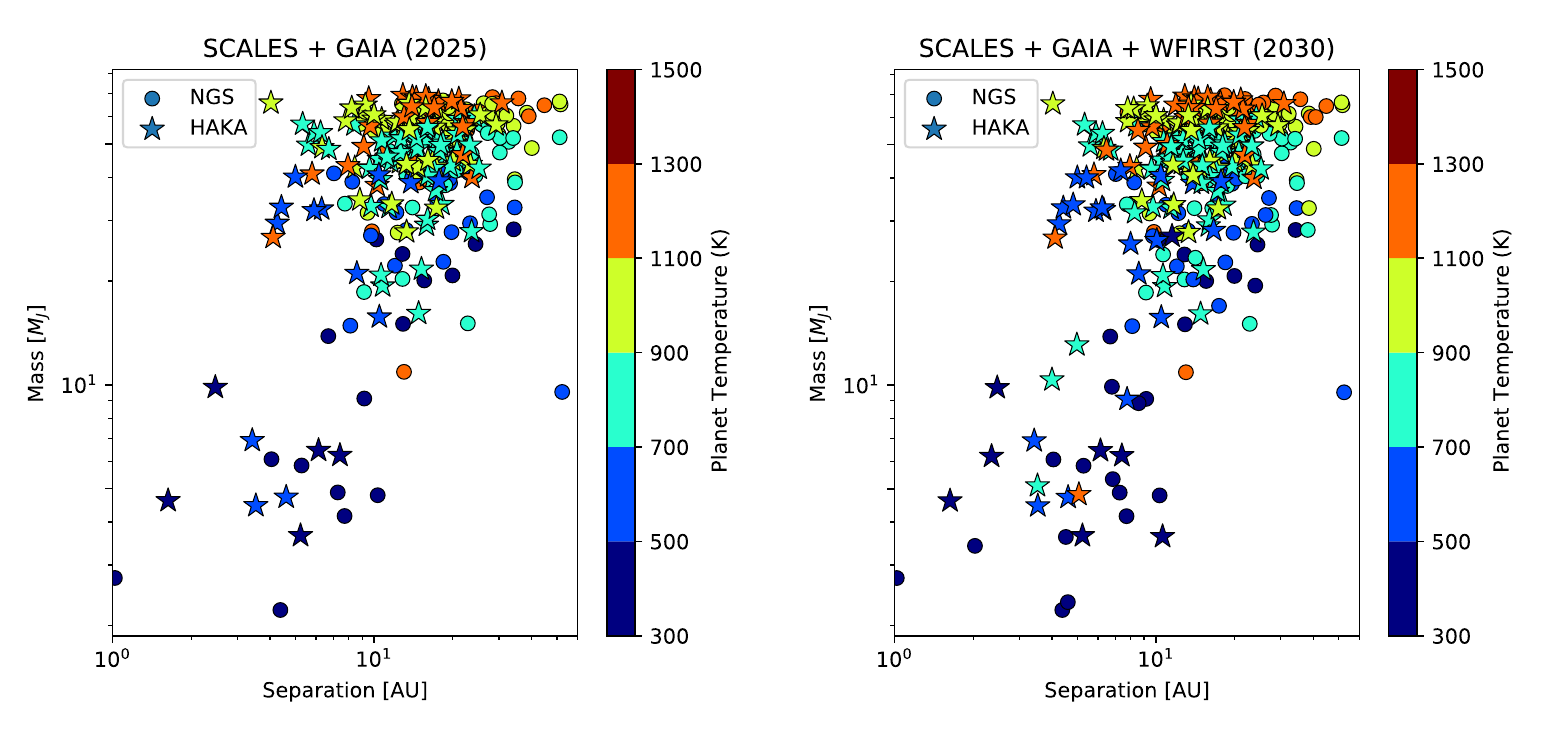}
    \caption{Scattered symbols show planets identified by Gaia (left) and Gaia plus WFIRST astrometry (right) that would be detectable by SCALES using angular differential imaging with 10 hours of total integration and $\sim80^\circ$ of parallactic angle evolution. The circles show planets within the contrast limits of Keck's current natural guide star (NGS) adaptive optics system. The stars indicate planets with contrasts that would require the planned High-Order All Sky Keck Adaptive Optics (HAKA) upgrade, which is expected to be operational when SCALES comes online.}
    \label{fig:gaia}
\end{figure}

\begin{figure}
    \centering
    \includegraphics[width=\textwidth]{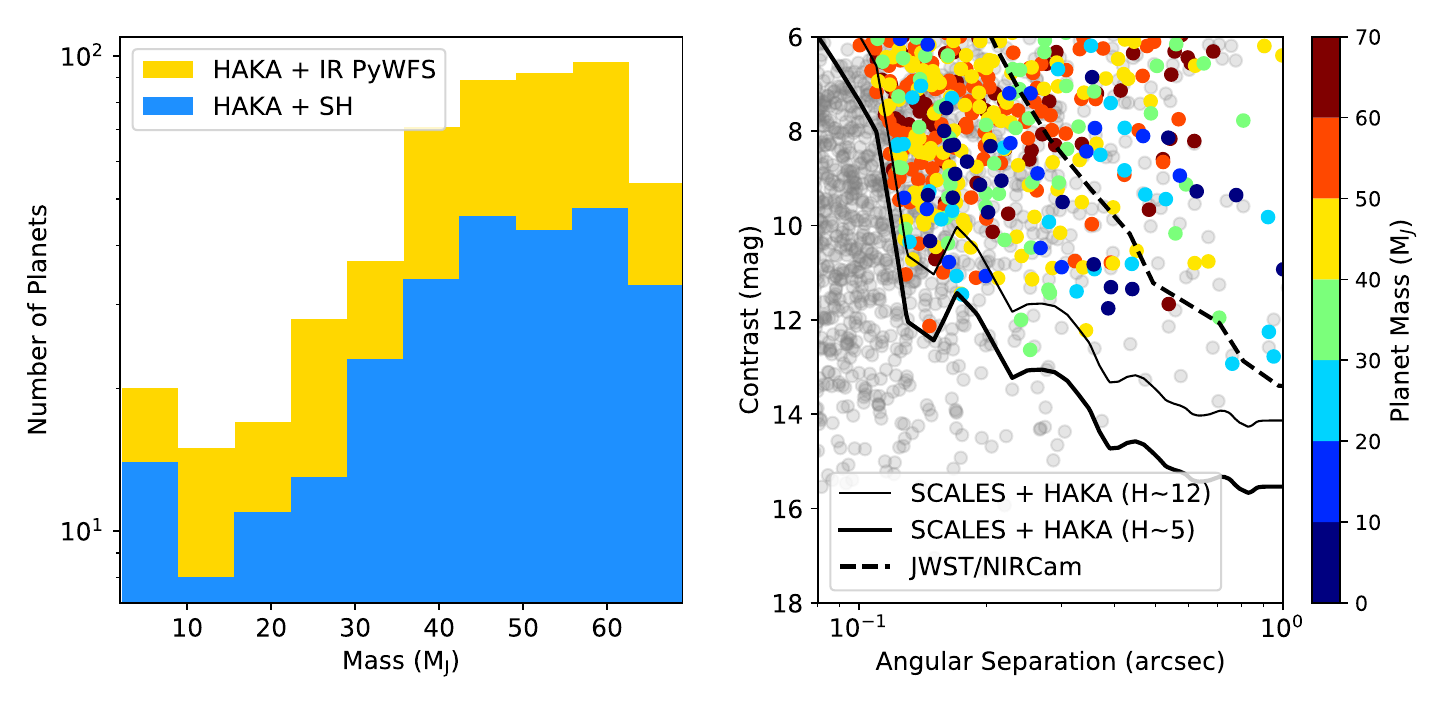}
    \caption{Left: Histograms show the boost in detectable Gaia planets and brown dwarfs for HAKA with an H-band Pyramid wavefront sensor (IR PyWFS; yellow), compared to HAKA with an R-band Shack-Hartmann wavefront sensor (blue). Right: Scattered grey points show simulated Gaia-detected exoplanets following the methodology described Section \ref{sec:gaia}. Colored points show planets detectable with 10 hours of SCALES integration and $\sim80^\circ$ of parallactic angle evolution. The thick and thin black lines show the anticipated HAKA contrast curve with the IR PyWFS for bright and faint guide stars, respectively. The thick dashed line shows the observed \textit{JWST}/NIRCam contrast curve from Carter et al.~2023.\cite{2023ApJ...951L..20C}}
    \label{fig:irhaka_jwst}
\end{figure}

Figure \ref{fig:gaia} shows the resulting planet yields. 
With current Keck NGS adaptive optics, SCALES would be capable of detecting 9 planets after Gaia's extended mission concludes in 2025, and 14 planets after an epoch of WFIRST astrometry in 2030.
Upgrading to the HAKA deformable mirror increases the expected yields to 17 and 28 planets for Gaia and Gaia + WFIRST, respectively. 
Figure \ref{fig:irhaka_jwst} shows the impact of upgrading the HAKA system to include an H-band infrared pyramid wavefront sensor (IR PyWFS),\cite{2020JATIS...6c9003B} and places these planet yields in the context of \textit{JWST}.
Due to the cool temperatures of the Gaia host stars with planets detectable by SCALES, wavefront sensing at H band significantly increases the expected number of planet detections (to 25 and 46 for Gaia and Gaia + WFIRST, respectively). 

As shown in Figure \ref{fig:irhaka_jwst}, many of these planets are uniquely detectable by SCALES, even in the era of \textit{JWST}.
The larger Keck diameter and expected adaptive optics performance provide SCALES with a tighter inner working angle than e.g.~NIRCam coronagraphy.\cite{2023ApJ...951L..20C} 
\textit{JWST's} sensitivity will outperform SCALES, meaning it will access cold, wide-separation planets that SCALES cannot (grey points above dashed line in Figure \ref{fig:irhaka_jwst}). 
However, the majority of the simulated Gaia planet population lies at tighter angular separations that require the SCALES inner working angle for detection. 
SCALES and \textit{JWST} will thus probe complementary regions of the Gaia planet parameter space.

\subsection{Atmospheric Abundances at Medium Spectral Resolution}

SCALES' medium spectral resolution mode is designed to enable robust (log abundance uncertainties $<$ 0.1) measurements of molecules such as CH$_4$, NH$_3$, and H$_2$O, as well as effective temperature measurements with precision $< 10$ K (Figure \ref{fig:traceability}). 
For hotter detectable planets with masses $\gtrsim5~\mathrm{M_J}$, SCALES will also be sensitive to additional molecules such as CO, CO$_2$, and PH$_3$, with expected log abundance uncertainties of $\lesssim 1$.

\begin{figure}
    \centering
    \includegraphics[width=\textwidth]{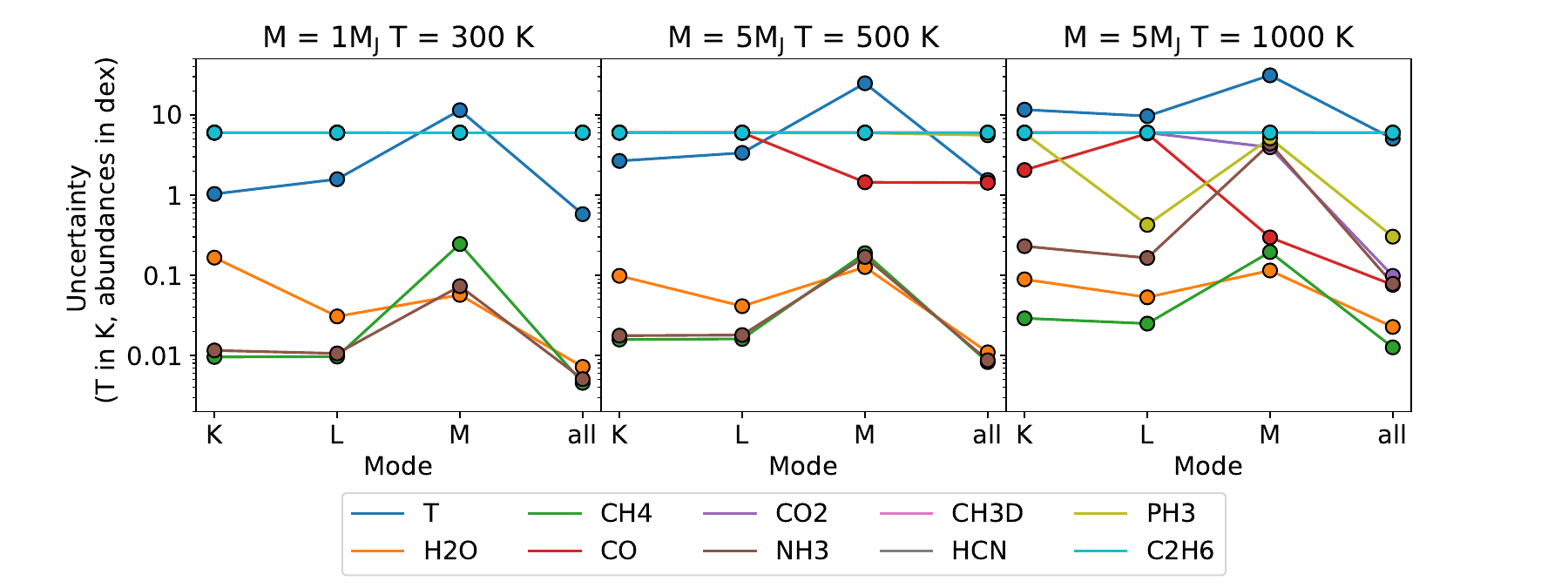}
    \caption{Sensitivity analysis for SCALES' medium resolution K, L, and M band modes, and combined datasets from all three modes, for three solar metallicity planets observed with a signal to noise ratio of $\sim3$. Each panel shows the uncertainty in abundance and temperature measurements as a function of the medium-spectral-resolution observing bandpass.  Left: Planet mass = 1 $\mathrm{M_J}$, radius = 1 R$_\mathrm{J}$, temperature = 300 K. Middle: Planet mass = 5 $\mathrm{M_J}$, radius = 1 R$_\mathrm{J}$, temperature = 500 K. Right: Planet mass = 5 $\mathrm{M_J}$, radius = 1 R$_\mathrm{J}$, temperature = 1000 K.}
    \label{fig:atm_sims}
\end{figure}

We follow the procedure from Batalha et al.~(2017)\cite{2017AJ....153..151B} and Batalha et al. ~(2018)\cite{2018ApJ...856L..34B}, to use Jacobians to demonstrate SCALES' sensitivity to changes in model atmospheric parameters. 
We generate simple error covariance matrices describing SCALES' precision at each wavelength channel, and use these along with the Jacobians to calculate the uncertainties of each model parameter after a measurement. 
We explore error covariance matrices corresponding to signal to noise ratios ranging from 1 - 15 for planets with temperatures of 300 K to 1000 K, and masses of 1 to 5 Jupiter masses.
We show the results for a signal to noise ratio of 3 in Figure \ref{fig:atm_sims}.

Figure \ref{fig:atm_sims} shows that a signal to noise ratio of 3 in SCALES' L and M band medium resolution modes is sufficient to meet the top level science goal in Figure \ref{fig:traceability}. 
This corresponds to a sensitivity per wavelength bin of $\sim18$ mag at L band, and $\sim15$ mag at M band. 
We use \texttt{scalessim} to demonstrate that we can achieve this level of sensitivity by generating medium resolution observations of cases shown in Figure \ref{fig:atm_sims}, observed at a distance of 15 pc.  
We assume 25 hours of total integration for a T = 300 K planet, and 10 hours of total integration for a T = 500 K planet. 

\begin{figure}[htp]
    \centering
    \includegraphics[width=\textwidth]{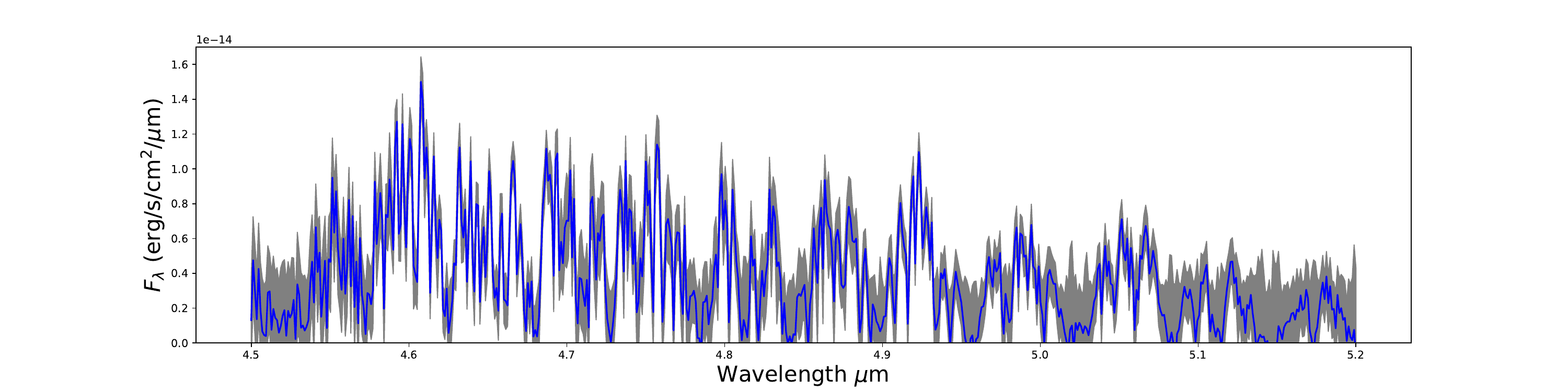}
    \caption[Simulated Gaia planet spectroscopy with SCALES.]{
        Simulated SCALES medium resolution M band observations of a 300 K, 1 R$_\mathrm{J}$ planet at a distance of 15 pc. The assumed integration time is 25 hours, and the extracted spectrum is shown in blue with error bars overlaid in grey. 
    }
    \label{fig:gaia1}
\end{figure}

\begin{figure}[htp]
    \centering
    \includegraphics[width=\textwidth]{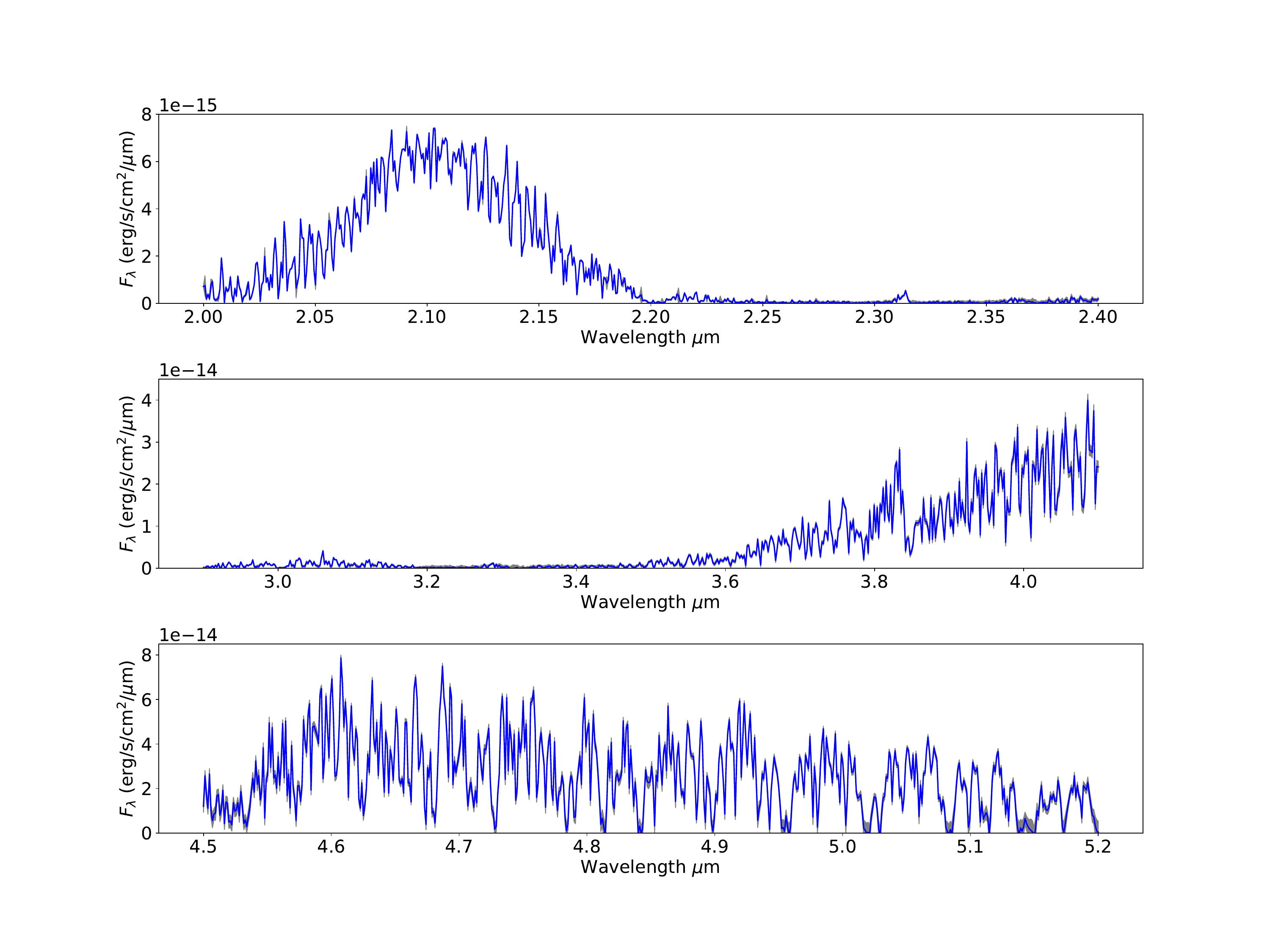}
    \caption[Simulated Gaia planet spectroscopy with SCALES.]{
        Simulated SCALES medium resolution K band (top), L band (middle), and M band (bottom) observations of a 500 K, 1 R$_\mathrm{J}$ planet at a distance of 15 pc. The assumed integration time is 10 hours, and the extracted spectrum is shown in blue with error bars overlaid in grey. 
    }
    \label{fig:gaia4}
\end{figure}

Figures \ref{fig:gaia1} and \ref{fig:gaia4} show the results. 
We can detect the warmer planet in all three bands, with signal to noise ratios varying from a few to a few tens. 
We detect the T = 300 K planet in M band only, since it is beyond the sensitivity limits of K and L bands. 
Despite these non-detections, Figure \ref{fig:atm_sims} shows that this is sufficient to constrain CH$_4$, NH$_3$, and H$_2$O in the cold planet to better than 0.2 dex. 
We note that perfect telluric correction is again assumed here. 
In future work we will include the variable precipitable water vapor described in Martinez et al.~(2023)\cite{martinez_inproc}, which will be particularly relevant for these atmospheric abundance measurements.

\subsection{Protoplanet Detection and Characterization}

Thanks to advances in adaptive optics and post-processing techniques, we have begun to directly detect and characterize actively forming planets embedded in protoplanetary disks.\cite{2018A&A...617A..44K,2019NatAs...3..749H} 
Furthermore, broadband direct imaging studies of young star systems have demonstrated the importance of spectrally-dispersed observations for protoplanet confirmation.
Circumstellar dust has been shown to complicate forming planet searches, with disk features sometimes masquerading as protoplanet companions\cite{2017AJ....153..264F,2023ApJ...953...55S} or making it difficult to distinguish true companions from the disk surroundings.\cite{2019NatAs...3..749H} 

One strategy to overcome these difficulties is to search for protoplanets in hydrogen line emission, since shocked gas would emit strongly in hydrogen lines as it falls into the potential well of the forming planet.\cite{2020arXiv201106608A}
This technique led to the identification of PDS 70 c,\cite{2019NatAs...3..749H}  which, when observed at K and L bands appears to be part of a circumstellar disk rim.\cite{2018A&A...617A..44K}
In addition to hydrogen lines, spectral slopes can be used to distinguish self-luminous protoplanet emission from forward scattered light by dust, since protoplanet continua rise as wavelength increases\cite{2015ApJ...803L...4E} while dust opacities fall off. 
The SCALES precursor, ALES, recently demonstrated this by identifying a protoplanet driving the spiral arms in MWC 758.\cite{2023NatAs.tmp..146W}
This protoplanet, MWC 758 c, has a significantly different spectral slope from its disk surroundings.

\begin{figure}
    \centering
    \includegraphics{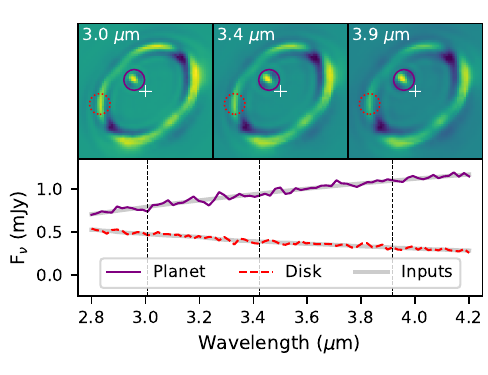}
    \caption{Example processed SCALES observations of a PDS 70 analog. Top panels show slices of a simulated SCALES L band observation of a PDS 70-like planet in a gapped protoplanetary disk. Bottom panels show the injected (grey lines) and recovered spectra at the protoplanet position (purple solid line) and in one region of the disk (dashed red line). The vertical lines in the bottom panel indicate the wavelengths of the three images in the top panels. SCALES will distinguish between self-luminous protoplanet signals and forward scattered light by disk material via their differing spectral slopes. }
    \label{fig:pds70}
\end{figure}

SCALES low-resolution spectroscopy from $2-5~\mu$m will be capable of characterizing protoplanet spectral slopes while distinguishing them from surrounding protoplanetary disk material.
Figure \ref{fig:pds70} illustrates this for a PDS 70 analog system observed with SCALES' L band low-resolution IFS mode. 
Comparing extracted spectra for the disk rim and embedded planet shows that SCALES will be sensitive to the differences in spectral slope between protoplanets and forward scattered light. 
The measured protoplanet continuua can then be used to constrain planet mass times accretion rate.\cite{2015ApJ...803L...4E}

SCALES' medium resolution modes will also provide access to hydrogen lines such as Br$\gamma$ ($2.166~\mu$m) and Br$\alpha$ ($4.05~\mu$m), which can also be used to constrain planet mass times accretion rate. 
These lines will further disentangle self-luminous protoplanets from forward scattered light by dust, and would suffer less from extinction than the more typically used H$\alpha$ line ($0.6563~\mu$m). 
See Martinez et al.~(2023)\cite{martinez_inproc} in these proceedings for a discussion of SCALES protoplanet characterization and potential specialized medium-resolution upgrades for tracing planetary accretion geometries.\cite{martinez_inproc}

\subsection{Protoplanetary Disk Mapping}

SCALES will be capable of spatially and spectrally resolving different dust and gas components in protoplanetary disks.
As shown in Figure \ref{fig:traceability}, SCALES' top level science goals for this case are to spatially resolve 
(1) dust density variations of $\sim10\%$ on scales of $\sim$ 10 AU around nearby T Tauri and Herbig Ae/Be stars, 
(2) H$_2$O ice features, with sensitivity to $\sim10\%$ abundance by mass on scales of $\sim$ 10 AU around nearby T Tauri stars, 
and
(3) PAH features, with sensitivity to $\sim10\%$ abundance by mass on scales of $\sim$ 10 AU around nearby Herbig Ae/Be stars. 

\begin{figure}
    \centering
    \includegraphics[width=6.5in]{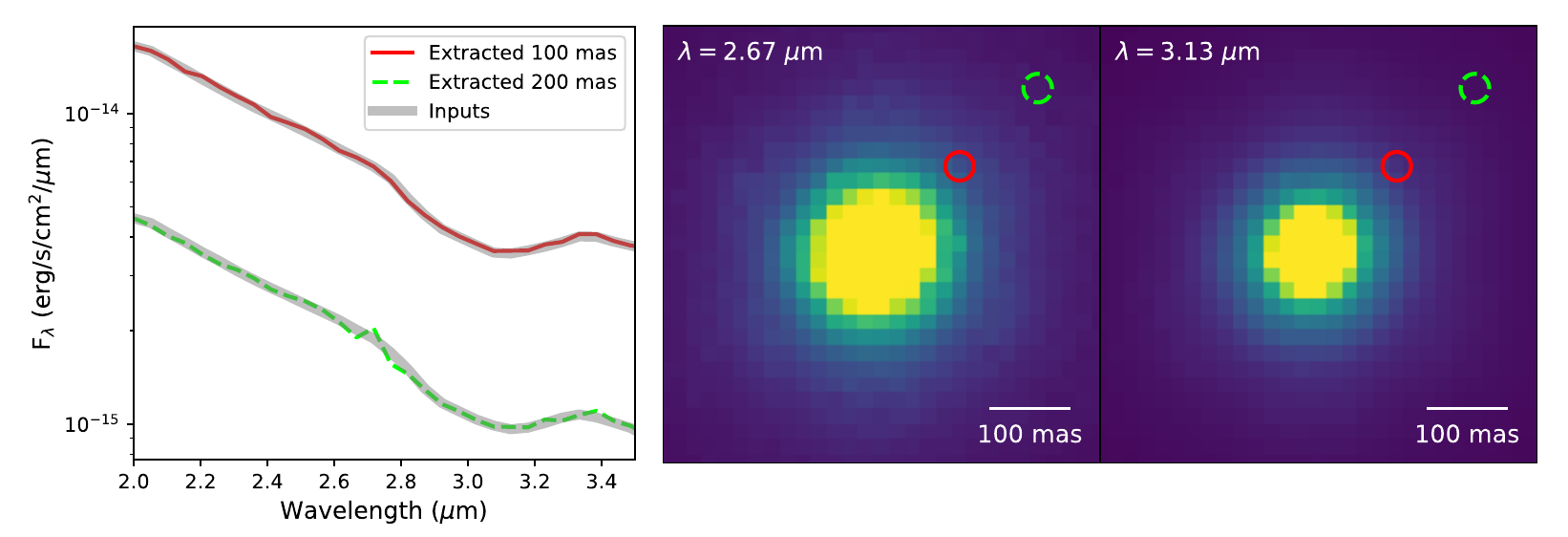}
    \caption{Simulated SCALES observations of a protoplanetary disk with 10\% water ice by mass. The left panel shows spectra extracted from \texttt{scalessim} mock observations of a water-ice-rich protoplanetary disk, measured for a single angular resolution element at two locations in the disk. These locations are indicated by the corresponding circles in the two simulated images (right panels). SCALES' technical requirements are set so that it will be sensitive to these absorption features at a wide range of protoplanetary disk radii.}
    \label{fig:h2o_locs}
\end{figure}

To track the technical requirements for protoplanetary disk mapping, we simulate radiative transfer images of protoplanetary disks using \texttt{pdspy} and \texttt{RADMC-3D}\cite{patrick_sheehan_2018_2455079, 2012ascl.soft02015D} and use \texttt{scalessim} to simulate SCALES datacubes. 
Figure \ref{fig:h2o_locs} shows an example  for water ice mapping, with absorption features comparable to those seen in previous imaging studies using dedicated filters.\cite{2009ApJ...690L.110H}.
SCALES' sensitivity is high enough to robustly detect H$_2$O absorption features both close to the central star where the disk is bright and far from the star where the disk is faint (Figure \ref{fig:h2o_locs}).

\subsection{Solar System Monitoring}

Infrared temporal monitoring of Solar System objects at Keck has provided a unique view of their surface and atmospheric phenomena \cite{2016Icar..280..405D,2018DPS....5011413A}. 
These include, but are not limited to, eruptions on the surface of Jupiter's moon Io, weather patterns and storms on Uranus, Neptune, and Titan, and seasonal variations in haze levels on planets and giant planet moons. 
Much of this work has been done with imaging during twilight time,\cite{2018DPS....5011413A} and SCALES will improve upon it by providing spatially-resolved spectroscopy. 
We set the top-level SCALES science objectives for Solar System objects in this context, focusing on Io volcanism and weather patterns on Uranus, Neptune, and Titan. 
These are: (1) to detect and monitor volcanoes on Io with sizes of $\sim300$ km and temperatures down to 300 K, and to constrain their temperatures to within $\sim30$ K, and (2) to map storms and methane features on Uranus, Neptune, and Titan with similar spatial resolution.

\begin{figure}
    \centering
    \includegraphics{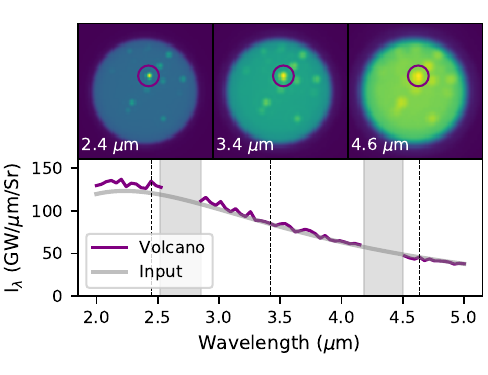}
    \caption{Simulated SCALES low-resolution SED mode observations of volcanism on Io. The top panels show images at three wavelengths in SCALES' SED mode, indicated by the vertical lines in the bottom panel. The bottom panel shows the input spectrum for the volcano circled in the top panels (grey curve), along the with extracted spectrum (purple curve). The grey shaded regions show regions of wavelength space where the sky transmission is very low.}
    \label{fig:io_sim}
\end{figure}

Figure \ref{fig:io_sim} shows an example simulated SCALES observation of Io volcanoes, assuming blackbody volcano spectra with a range of temperatures set by de Kleer et al.~(2016) ($\sim300-1200$ K).\cite{2016Icar..280..378D}
The spectral resolution and coverage of SCALES' low-resolution 2-5 $\mu$m SED mode will enable more robust constraints on volcano temperatures than broadband imaging (Figure \ref{fig:io_sim}, bottom panel). 
SCALES' R$\sim$200 K band and R$\sim$250 CH$_4$ low-resolution modes will also provide the required wavelength coverage and spectral resolution to measure methane features at 
 $\sim2-2.25~\mu$m and $\sim3-3.5~\mu$m on Uranus, Neptune, and Titan.\cite{2011Icar..213..218B,2005EM&P...96..119B}

\subsection{Non-(Exo)planetary Science Applications}

While the primary science drivers for the SCALES IFS are the cases outlined above, the SCALES team tracks requirements for non-(exo)planetary science cases as well. 
The resolving power and wavelength coverage of the medium-resolution M band mode will make it possible to characterize the compositions of knots in supernova remnants.\cite{1995ApJ...440..706R,2001ApJS..133..161F}
In addition to mapping PAH emission around Herbig Ae/Be stars, the low-resolution PAH mode will enable mapping of these features in nearby active galactic nuclei tori.\cite{2003A&A...398..101M}
SCALES low-resolution IFS modes will also enable astrometric and spectroscopic characterization of stars and dust-shrouded objects (G sources)\cite{2020Natur.577..337C} orbiting Sagittarius A*.

\section{PREPARATIONS FOR EARLY SCALES SCIENCE}
The technical capabilities offered by SCALES will lead to new progress in exoplanet detection and characterization, a wide range of broader (exo)planetary studies, and a number of galactic and extragalactic science cases. 
Shortly after it sees first light in 2025, the SCALES Science Team will begin a Science Demonstration Survey to produce proof-of-concept observations for SCALES' representative science applications. 
The target selection for this survey will be driven by community input, and the reduced data products will be made publicly available on the Keck Observatory Archive. 
Preparations for this program are currently underway, and we encourage interested readers to reach out to S.S. and use \texttt{scalessim} to explore potential SCALES outcomes for their science.

\acknowledgments 
We are grateful to the Heising-Simons Foundation, the Alfred P. Sloan Foundation, and the Mt. Cuba Astronomical Foundation for their generous support of our efforts. This project also benefited from work conducted under the NSF Graduate Research Fellowship Program. S.S. is supported by the National Science Foundation under MRI Grant No. 2216481. R.A.M is supported by the National Science Foundation MPS-Ascend Postdoctoral Research Fellowship under Grant No.~2213312.

\bibliography{references2} 
\bibliographystyle{spiebib} 

\end{document}